\documentclass{icrc2009}

\usepackage{graphicx}   
\usepackage[caption=false]{caption}    
\usepackage[font=footnotesize]{subfig} 
\usepackage{fixltx2e}
\usepackage{url}

\newcommand{\shorttitle}[1]%
{\markboth{Proceedings of the 31\MakeLowercase{$^{st}$} ICRC, {\L}\'{o}d\'{z} 2009}{#1} }
\newcommand{\etal}{\MakeLowercase{\textit{et al. }}} 

\newcommand{\unit}[1]{\ensuremath{\, \mathrm{#1}}}
\newcommand{\e}[1]{\ensuremath{\cdot 10^{#1}}}

\begin{document}
\title{Reconstruction of IceCube coincident events and study of composition-sensitive observables using both the surface and deep detector}

\author{\IEEEauthorblockN{Tom Feusels\IEEEauthorrefmark{1},
                          Jonathan Eisch\IEEEauthorrefmark{2} and
			  Chen Xu\IEEEauthorrefmark{3},
                          for the IceCube Collaboration\IEEEauthorrefmark{4}}
                            \\
\IEEEauthorblockA{\IEEEauthorrefmark{1}Dept. of Subatomic and Radiation Physics, University of Gent, B-9000 Gent, Belgium}
\IEEEauthorblockA{\IEEEauthorrefmark{2}Dept. of Physics, University of Wisconsin, Madison, WI 53706, USA}
\IEEEauthorblockA{\IEEEauthorrefmark{3}Bartol Research Institute, University of Delaware, Newark, DE 19716, USA}
\IEEEauthorblockA{\IEEEauthorrefmark{4}See the special section in these proceedings}}

\shorttitle{T. Feusels \etal Reconstruction of IceCube coincident events}
\maketitle

\begin{abstract}
The combined information from cosmic ray air showers that trigger both the surface and underground parts of the IceCube Neutrino Observatory allows the reconstruction of both the energy and mass of the primary particle through the knee region of the energy spectrum and above. The properties of high-energy muon bundles, created early in the formation of extensive air showers and capable of penetrating deep into the ice, are related to the primary energy and composition.

New methods for reconstructing the direction and composition-sensitive properties of muon bundles are shown. Based on a likelihood minimization procedure using IceCube signals, and accounting for photon propagation, ice properties, and the energy loss processes of muons in ice, the muon bundle energy loss is reconstructed. The results of the high-energy muon bundle reconstruction in the deep ice and the reconstruction of the lateral distribution of low energy particles in the surface detector can be combined to study primary composition and energy.  The performance and composition sensitivity for both simulated and experimental data are discussed. 
\end{abstract}

 
\section{Introduction}
The cosmic ray spectrum covers many orders of magnitude in both energy and flux. In the energy range accessible to the IceCube Neutrino Observatory ($\sim$0.3\unit{PeV} to 1\unit{EeV}) the slope of the spectrum remains mostly constant, except for a feature at around 3\unit{PeV} where the spectrum steepens.  This feature is called the knee of the cosmic ray spectrum, and its origin is unknown.  Proposed explanations include changes in acceleration mechanisms or cosmic rays leaking from the galactic magnetic field starting at this energy. Measurement of mass composition in this range could give clues to the origin of these cosmic rays.  The IceCube detector, together with the IceTop air shower detector, provides an opportunity to measure the composition of cosmic ray particles in the region of the knee and beyond.

The IceTop detector, high on the Antarctic Plain at an average atmospheric depth of 680\unit{g/cm^{2}}, consists of a hexagonal grid of detector stations 125\unit{m} apart.  Each station consists of two ice tanks that act as Cherenkov media for measuring mainly the electromagnetic component of cosmic ray air showers. In each tank, two Digital Optical Modules (DOMs) are deployed, which contain a 10\unit{inch} PMT and digital readout and control electronics.  The primary energy can be reconstructed by the IceTop surface array \cite{kislat}. 

Deep below each IceTop station is a string of the IceCube detector with 60 DOMs evenly spaced between 1.5 and 2.5\unit{km} in the ice.  Combined, the IceTop and IceCube arrays can reconstruct the air shower core position and direction while measuring the shower signal strength at the surface and the energy deposition of the high-energy muon bundle in the deep ice.  With these measurements, the energy and mass of the primary cosmic rays can be reconstructed. 

A characteristic difference between showers induced by light and heavy nuclei for a fixed primary energy is their number of muons, with higher mass primaries producing more muon-rich showers.  However, the muon multiplicity is not directly measured by either the IceTop or IceCube detectors. The deep IceCube detector is sensitive to Cherenkov light coming mainly from energy loss processes of high-energy muon bundles. This energy loss is a convolution of the muon multiplicity of the shower, the muon energy distribution and the energy loss of a single muon.  If the energy loss behavior of muon bundles can be reconstructed accurately, it can be used as a primary mass indicator \cite{muonbundleEloss}. In Section \ref{sec:ereco}, the reconstructed muon bundle track described in Section \ref{sec:dirreco} will be used as a seed to reconstruct the muon bundle energy loss. Simulated data is compared to experimental data in Section \ref{sec:coinc} to examine the detector performance and its sensitivity to composition.

\section{Data Sample and Simulation}
\label{sec:simdatasamp}
Our experimental data sample was taken from the month of September, 2008.  At that time, the detector consisted of 40 IceTop stations and 40 IceCube strings. A total livetime of 28.47 days was obtained by selecting only runs where both detectors were stable. The events were processed at the South Pole with a filter which required at least three triggered IceTop stations and at least 8 triggered IceCube DOMs. After this filter about 3.31\e{6} coincident events remained.

To study the direction and energy reconstruction of muon bundles in the ice, a large number of proton and iron showers between 10\unit{TeV} and 46.5\unit{PeV} were simulated with the CORSIKA \cite{Corsika} package. SIBYLL 2.1 \cite{Sibyll, Sibyll2} was used as the high-energy (\textgreater 80\unit{TeV}) hadronic interaction model, while FLUKA08 \cite{Fluka1, Fluka2} was used as the low energy hadronic interaction model.  For this study, CORSIKA was configured to use a model of the South Pole atmosphere typical for the month of July \cite{Atmos}.  The showers were simulated according to an E$^{-1}$ spectrum and then reweighted according to an E$^{-2.7}$ spectrum before the knee (at 3\unit{PeV}) and an E$^{-3.0}$ spectrum after the knee. 

The IceCube software environment was used to resample each CORSIKA shower 500 times on and around the detector, to propagate the high-energy muons through the ice, and to simulate the detector response and trigger. The simulation was filtered the same way as data and yielded about 9.0\e{4} proton and 9.0\e{4} iron events.

\section{Direction Reconstruction}
\label{sec:dirreco}
The direction reconstruction by IceTop will be described first because it will be used later as a seed for an IceCube muon bundle reconstruction algorithm.

\begin{figure}[!t]
  \centering
  \includegraphics[width=2.5in,angle = 270]{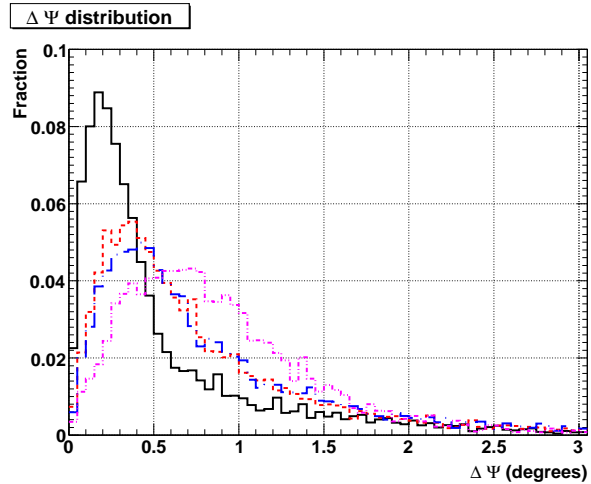} 
  \caption{The distribution of the angles between the true direction and the reconstructed direction for simulations of iron showers is shown for different algorithms. The dot-dot-dashed curve shows the reconstruction which uses IceTop information alone. For the dot-dashed line, the reconstructed core position by IceTop is fixed and the zenith and azimuth are determined by using a muon bundle algorithm seeded with the track determined by IceTop. The solid line illustrates the ideal limit of this reconstruction method by using the true shower core position instead of the reconstructed one. The dashed curve is obtained when a second iteration between IceTop and IceCube algorithms is used.}
  \label{resol}
 \end{figure}

\begin{figure}[!t]
  \centering
  \includegraphics[width=2.5in,angle = 270]{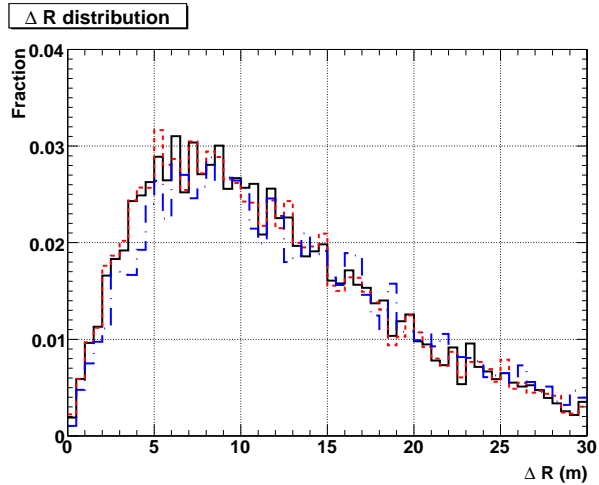} 
  \caption{The distribution of distances between the true shower core position and the reconstructed core position at the surface for iron showers is plotted for different reconstructions. For the dash-dotted line the IceTop DLP function that determines the core position used a first guess direction. The dashed line, where the IceTop algorithm was seeded with a better direction reconstruction from the deep ice, gives a slightly better core reconstruction. When the true direction is used as a seed, the ideal limit represented by the solid line is obtained.} 
  \label{resol_core}
 \end{figure}

An initial shower core position and direction is determined using the extracted times and charges of the recorded pulses from the IceTop tanks. The first guess core position is the calculated center of gravity of tank signals, while the initial direction reconstruction assumes a flat shower front. This shower core and direction are then used as a seed to fit the lateral distribution of pulses with the double logarithmic parabola (DLP) function described in \cite{lateralfit}. Because the reconstruction of the core and direction are highly correlated, the resolution is improved by fitting the shower core position and the shower direction together, with a curved shower front and the DLP lateral particle distribution. This fit, and some geometrical quality cuts discussed later, gives an angular resolution\footnote{The resolution of observable $Y$ is defined according to $\int_0^x p(\Delta Y)\, \mathrm{d}\Delta Y = 0.683$, where $x$ is the resolution and $p(\Delta Y)$ the frequency of distances between the true and the reconstructed observable.} of 1.0$^\circ$ for iron showers (see the dot-dot-dashed line in Fig. \ref{resol}) and a core resolution of 15.0\unit{m} (see dot-dashed line in Fig. \ref{resol_core}). 

The resolution depends strongly on where the shower core lands with respect to the IceTop array. This analysis uses only events with a shower core reconstructed within the geometrical area of the IceTop detector. The reconstructed muon bundle track was also required to pass within the instrumented volume of the IceCube detector.

The direction that was already reconstructed by the IceTop algorithms alone can serve as a seed for an algorithm more specialized in reconstructing muon bundles in the ice. This algorithm uses only the charges measured by the IceCube DOMs and takes into account the range-out of the muons in a bundle \cite{muonbundlerecoLDF}. By keeping the reconstructed core position on the surface fixed, a large lever arm of at least 1500\unit{m} is obtained. This limits the track parameters (zenith and azimuth) during the minimization procedure and reduces the number of free track parameters from 5 to 2.

In Fig. \ref{resol}, it can be seen that this method (dot-dashed line) improves on the angular resolution determined by IceTop alone. If the core position can be determined more accurately, the direction reconstruction will be even better. The solid line on Fig. \ref{resol} represents the ideal limit, obtained using the true core position. Therefore, to improve the core resolution the new direction is kept fixed and used to seed the IceTop lateral distribution function which then only fits the core position (dashed line on Fig. \ref{resol_core}). Iterating over both the surface and the deep detector reconstructions with progressively better core position and direction seeds, leads to the optimal resolution. The ideal limit for the core resolution is acquired by seeding the IceTop algorithm with the true direction and is shown on Fig. \ref{resol_core} (solid line). After the second iteration this limit is already obtained and gives a core resolution of 14.0\unit{m} and an angular resolution of 0.9$^\circ$ (see dashed lines in Figures \ref{resol} and \ref{resol_core}).

Using this combined method for muon bundle direction reconstruction, an almost energy independent angular resolution of 0.8$^\circ$ and a core resolution of 12.5\unit{m} is obtained for proton induced showers. An improvement of the core resolution and direction resolution also improves the resolution of the reconstructed shower size and shower age.

\section{Muon Bundle Energy Reconstruction}
\label{sec:ereco}
An energy reconstruction algorithm for single muons was previously reported in Ref. \cite{photorec}. Using IceCube signals and lookup tables together with a previously reconstructed track a constant energy loss is fit with a likelihood function. The lookup tables model the South Pole ice properties and the propagation of Cherenkov photons through the ice.  A single high-energy muon above 730~GeV loses energy mainly by radiative processes like Bremsstrahlung and pair production, which produce secondary electromagnetic cascades in the ice along the muon bundle track. Therefore, an infinite light source with mono-energetic cascades every meter is used as a model for a single muon. 

This light model can also be used for muon bundles. The main difference is that a slant depth dependent energy loss will be needed because of the range-out of muons.  Here, the slant depth is defined as the distance along the muon bundle track between the point where the muon enters the ice and the point where the Cherenkov light is emitted. 

The equation for the muon bundle energy loss is:
\begin{equation}
 \left( \frac{dE_{\mu}}{dX} \right)_{\mathrm{Bundle}} (X) = \int_{E_{\mathrm{min(surf)}}}^{E_{\mathrm{max(surf)}}} \frac{dN_{\mu}}{dE_{\mu}}\frac{dE_{\mu}}{dX} dE_{\mu}\mathrm{(surf)}\,,  \label{eq:muBu} 
\end{equation}
 where $\frac{dN_{\mu}}{dE_{\mu}} $ is the energy distribution of the muons and $\frac{dE_{\mu}}{dX}$ is the energy loss of a single muon. $E_{\mathrm{min(surf)}} = \frac{a}{b}\left(e^{bX}-1\right)$ is the minimum energy that a muon needs to get to depth $X$. $E_{\mathrm{max(surf)}} \propto \frac{E_0}{A}$ is the maximum energy a muon from a shower induced by a particle with $A$ nucleons and primary energy $E_0$ can have. 

Using a simple power-law as an approximation for the Elbert formula \cite{Elbert}, which describes the multiplicity of high-energy muons in air showers, the differential muon energy distribution becomes:

\begin{equation}
\frac{dN_{\mu}}{dE_{\mu}} = \gamma_{\mu}\, \kappa(A) \, \left(\frac{E_0}{A}\right)^{\gamma_{\mu}-1} E^{-\gamma_{\mu}-1}_\mu\,, \label{eq:Edistr}
\end{equation}
where $\gamma_{\mu} = 1.757$ is the muon integral spectral index and $\kappa$ is a normalization that depends on the shower properties.

With the solution of the single muon energy loss equation, $E_{\mu}(X) = \left(E_{\mu}(\mathrm{surf}) + \frac{a}{b}\right)\,e^{-bX}-\frac{a}{b}$, the average energy loss formula can be expressed as a function of the muon energy at the surface: 
\begin{eqnarray}
\frac{dE_{\mu}}{dX} (X) &=& - a - b E_{\mu}(X) \nonumber \\
                       &=& -b\left(E_{\mu}(\mathrm{surf}) +\frac{a}{b} \right) e^{-bX}\,, \label{eq:Eloss}
\end{eqnarray}
with $a$ = 0.260\unit{GeV/m}, the ionization energy loss constant and $b$ = 0.000357\unit{m}$^{-1}$, the stochastic energy loss constant from \cite{Eloss}.

The average muon bundle energy loss function is then obtained by integrating Eq. (\ref{eq:muBu}) using Eqs. (\ref{eq:Edistr}) and (\ref{eq:Eloss}). This energy loss fit function, with $\kappa$ and $E_0/A$ as free parameters, will be used to scale the expected charges in a DOM from the likelihood formula in \cite{photorec} instead of scaling it with a constant energy loss.

 \begin{figure}[!t]
  \centering
  \includegraphics[width=2.5in,angle = 270]{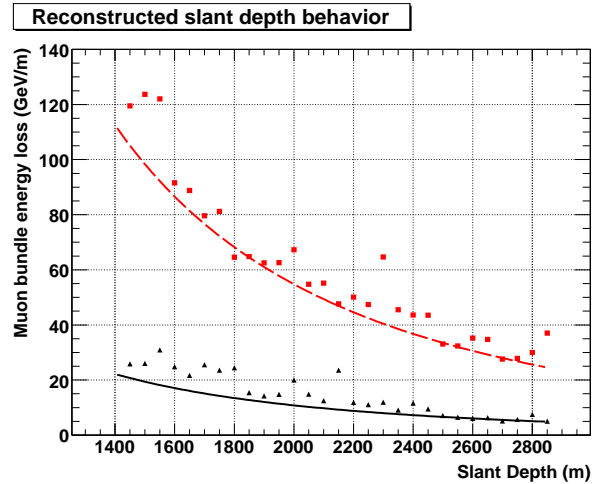} 
  \caption{An example of reconstructed muon bundle energy loss for a single 17~PeV iron and proton shower. The reconstructed slant depth behavior follows the true energy loss reasonably well. The triangles (squares) are the muon bundle energy loss processes for a certain proton (iron) 17\unit{PeV} shower with a zenith angle of 28.4$^\circ$ (12.6$^\circ$). The spread of the points illustrates the stochastic nature. The solid (dashed) line is the reconstructed muon bundle energy loss function, described in the text, for the proton (iron) shower.
  }
  \label{energy_resol}
 \end{figure}

In Fig. \ref{energy_resol}, the curves show the reconstructed muon bundle energy loss functions for 17\unit{PeV} primaries. The data points are the energy losses calculated from simulations. It can be clearly seen that the muon bundle energy loss function obtained by minimizing the likelihood function describes the depth behavior for these two showers better than a constant energy loss function. 

It has been shown in \cite{muonbundleEloss} that proton and iron showers are separated better by the muon bundle energy loss at smaller slant depths. The reconstructed IceCube composition-sensitive parameter which will be used further on in the coincidence analysis, is the energy loss at the top of the IceCube detector, at a slant depth of 1650\unit{m}. At this slant depth, Cherenkov light from showers with a zenith angle up to 30$^\circ$ can still be detected by the upper DOMs, making the energy reconstruction more accurate over the entire zenith range.

\section{Combined Primary Mass and Energy Reconstruction}
\label{sec:coinc}
 \begin{figure}[!t]
  \centering
  \includegraphics[width=2.5in,trim = 50 0 85 0]{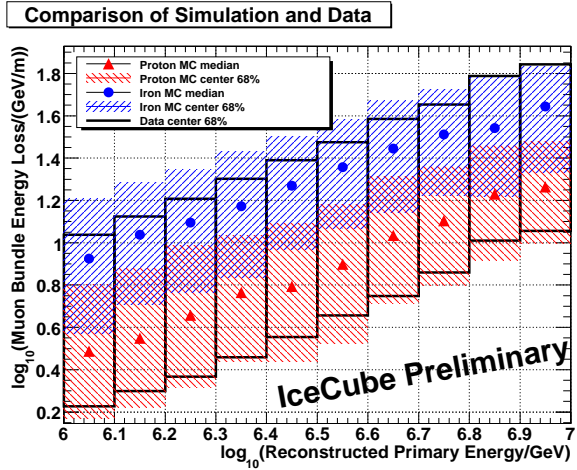}
  \caption{The reconstructed muon bundle energy loss evaluated at a slant depth of 1650\unit{m} versus the reconstructed shower primary energy.  Shown are the center 68\% of the proton simulation (right diagonal hashes), the iron simulation (left diagonal hashes) and the center 68\% of the data (enclosed in the rectangles).  The median of the distribution is also shown for the simulated data sets.  The events were filtered as described in Section \ref{sec:simdatasamp} with the additional quality requirement that the muon bundle reconstruction algorithm used signals from at least 50 DOMs, to remove events with poorly fit energy loss.  This quality cut removes a higher portion of muon bundles with low energy loss and will be accounted for in a full composition analysis. }
  \label{data_sim_comp}
 \end{figure}
 
 Fig. \ref{data_sim_comp} shows a comparison of proton and iron primary simulation and experimental data described in Section \ref{sec:simdatasamp} using the reconstruction methods described in the sections \ref{sec:dirreco} and \ref{sec:ereco}. While this plot has only rough quality cuts, it can be seen that the spread in the simulation is similar to the spread in the data.  The median muon bundle energy loss from an iron primary shower is approximately a factor of 2 higher in the ice than the median muon bundle energy loss from a proton primary shower for the same reconstructed primary energy.  There is a large overlap area, where shower to shower fluctuations overcome the effect of the primary mass.  It is difficult to reconstruct the primary mass of a single shower with any certainty due to these large fluctuations.

\section{Conclusion and Outlook}
\label{sec:concl}
The IceTop and IceCube detectors of the IceCube Neutrino Observatory can be used together for an improved air shower core location and direction reconstruction. The promising method of reconstructing the muon bundle energy loss behavior will be further developed to be used in a measurement of cosmic ray mass and energy.  This study used only a subsample of the available data; with more statistics and an enlarged detector these methods can be extended up to 1\unit{EeV}.  

\section{Acknowledgements}
This work is supported by the Office of Polar Programs of the National Science Foundation and by FWO-Flanders, Belgium.


\begin{thebibliography}{99}
   \bibitem{kislat} F.~Kislat \textit{et al.}, \emph{A First All-Particle Cosmic Ray Energy Spectrum From IceTop}, These Proceedings. 
  \bibitem{muonbundleEloss} X.~Bai \textit{et al.}, 
         \emph{Muon Bundle Energy Loss in Deep Underground Detector}, These Proceedings. 

   \bibitem{Corsika} D.~Heck \textit{et al.}, 
	            {\em CORSIKA FZKA 6019}, Forschungszentrum Karlsruhe, 1998.
    \bibitem{Sibyll} R.~S. Fletcher \textit{et al.}, 
	{\em SIBYLL: An event generator for simulation of high energy cosmic ray cascades}, Phys. Rev. D, {\bf 50} 5710, 1994.	
    \bibitem{Sibyll2} R.~Engel \textit{et al.}, 
	{\em Air shower calculations with the new version of SIBYLL}, In Proc. 26th ICRC, Salt Lake City, 1999.	
    \bibitem{Fluka1} A.~Fass\`{o} \textit{et al.}, 
	{\em FLUKA: a multi-particle transport code },
	CERN-2005-10 (2005), INFN/TC\_05/11, SLAC-R-773.
   \bibitem{Fluka2} G. ~Battistoni \textit{et al.},
	{\em The FLUKA code: Description and benchmarking },
	Proceedings of the Hadronic Shower Simulation Workshop 2006,
	Fermilab 6--8 September 2006, M. Albrow, R. Raja eds.,
	AIP Conference Proceeding 896, 31-49, (2007).
     \bibitem{Atmos} D.~Chirkin, Parameterization based on the MSIS-90-E model, 1997, Private communication. 

   \bibitem{lateralfit}  S.~Klepser \textit{et al.}, \emph{Lateral Distribution of Air Shower Signals and Initial Energy Spectrum above 1~PeV from IceTop},
                     In Proc. 30th ICRC, Merida, Mexico, 2007.
   \bibitem{muonbundlerecoLDF} K.~Rawlins, Ph.D. Dissertation, UW-Madison (2001).
   \bibitem{Eloss} P.~Mio\v{c}inovi\'{c}, Ph.D. Dissertation, UC Berkeley (2001).
   \bibitem{photorec}  S.~Grullon \textit{et al.}, \emph{Reconstruction of high-energy muon events in IceCube using waveforms},
                     In Proc. 30th ICRC, Merida, Mexico, 2007.
    \bibitem{Elbert} J.W.~Elbert, In Proc. DUMAND Summer Workshop (ed. A. Roberts), 1978, vol 2, p.101.
    
  \end{thebibliography}
\end{document}